\documentclass[aps,twocolumn,showpacs,prb,superscriptaddress,floatfix,longbibliography]{revtex4-2}

\usepackage{amssymb,amsmath,amsthm,bbm,bbold}
\usepackage{graphicx}
\usepackage{float}
\usepackage[english]{babel}
\usepackage{array}
\usepackage[T1]{fontenc}
\usepackage[usenames,dvipsnames]{xcolor}
\usepackage{hyperref}

\usepackage{comment}

\newcommand{\Tr}{\mathrm{Tr}}

\newcommand{\HN}{\mathcal{H}^{(N)}}
\newcommand{\be}{\begin{equation}}
\newcommand{\ee}{\end{equation}}

\def\B{ {\mathcal B} }

\def\F{ {\mathcal F} }

\newcommand{\Ho}{\mathcal{H}^{(1)}}
\newcommand{\Hl}{\mathcal{H}_l^{(1)}}
\newcommand{\Hs}{\mathcal{H}_s^{(1)}}
\newcommand{\HH}{\mathcal{H}}
\newcommand{\BH}{\mathcal{B}(\HH)}
\newcommand{\BHA}{\mathcal{B}(\HH_A)}
\newcommand{\BHB}{\mathcal{B}(\HH_B)}
\newcommand{\BHAB}{\mathcal{B}(\HH_{AB})}

\newcommand{\DD}{\mathcal{D}}

\newcommand{\CC}{\mathbbm{C}}

\def\Tr{{\rm{Tr}}}

\newcommand{\bra}[1]{\mbox{$\langle #1 |$}}
\newcommand{\ket}[1]{\mbox{$| #1 \rangle$}}

\begin{document}

\title{What Can Quantum Information Theory Offer to Quantum Chemistry?}

\author{Damiano Aliverti-Piuri}
\affiliation{Department of Physics, Arnold Sommerfeld Center for Theoretical Physics,
Ludwig-Maximilians-Universit\"at M\"unchen, Theresienstrasse 37, 80333 M\" unchen, Germany}
\affiliation{Munich Center for Quantum Science and Technology (MCQST), Schellingstrasse 4, 80799 M\"unchen, Germany}
\author{Kaustav Chatterjee}
\affiliation{Department of Physics, Arnold Sommerfeld Center for Theoretical Physics,
Ludwig-Maximilians-Universit\"at M\"unchen, Theresienstrasse 37, 80333 M\" unchen, Germany}
\affiliation{Munich Center for Quantum Science and Technology (MCQST), Schellingstrasse 4, 80799 M\"unchen, Germany}

\author{Lexin Ding}
\affiliation{Department of Physics, Arnold Sommerfeld Center for Theoretical Physics,
Ludwig-Maximilians-Universit\"at M\"unchen, Theresienstrasse 37, 80333 M\" unchen, Germany}
\affiliation{Munich Center for Quantum Science and Technology (MCQST), Schellingstrasse 4, 80799 M\"unchen, Germany}

\author{Ke Liao}
\affiliation{Department of Physics, Arnold Sommerfeld Center for Theoretical Physics,
Ludwig-Maximilians-Universit\"at M\"unchen, Theresienstrasse 37, 80333 M\" unchen, Germany}
\affiliation{Munich Center for Quantum Science and Technology (MCQST), Schellingstrasse 4, 80799 M\"unchen, Germany}

\author{Julia Liebert}
\affiliation{Department of Physics, Arnold Sommerfeld Center for Theoretical Physics,
Ludwig-Maximilians-Universit\"at M\"unchen, Theresienstrasse 37, 80333 M\" unchen, Germany}
\affiliation{Munich Center for Quantum Science and Technology (MCQST), Schellingstrasse 4, 80799 M\"unchen, Germany}

\author{Christian Schilling}
\email{c.schilling@physik.uni-muenchen.de}
\affiliation{Department of Physics, Arnold Sommerfeld Center for Theoretical Physics,
Ludwig-Maximilians-Universit\"at M\"unchen, Theresienstrasse 37, 80333 M\" unchen, Germany}
\affiliation{Munich Center for Quantum Science and Technology (MCQST), Schellingstrasse 4, 80799 M\"unchen, Germany}

\date{\today}

\begin{abstract}
It is the ultimate goal of this work to foster synergy between quantum chemistry and the flourishing field of quantum information theory. For this, we first translate quantum information concepts such as entanglement and correlation into the context of quantum chemical systems. In particular, we establish two conceptually distinct perspectives on `electron correlation' leading to a notion of orbital and particle correlation. We then demonstrate that particle correlation equals total orbital correlation minimized over all orbital bases. Accordingly, particle correlation resembles the minimal, thus intrinsic, complexity of many-electron wave functions while orbital correlation quantifies their complexity relative to a basis. We illustrate these concepts of intrinsic and extrinsic correlation complexity in molecular systems, which also manifests the crucial link between the two correlation pictures. Our results provide theoretical justification for the long-favored natural orbitals for simplifying electronic structures, and open new pathways for developing more efficient approaches towards the electron correlation problem.
\end{abstract}

\maketitle

\section{Introduction}\label{sec:intro}

Quantum chemistry (QChem) and its ability to accurately predict properties of molecules and materials is nowadays indispensable for a broad spectrum of modern quantum science. For instance, it deepens our understanding of chemical processes~\cite{almlof1974,dzubak2012a,kurashige2013b,sharma2014a,limanni2018,larsson2022}, as well as drives forward materials science~\cite{misawa2014a,Schimka2010,Booth2013,yang2014,gruber2018a,liao2019,zhang2019,cui2022,bogdanov2022,cui2023}.
The success of QChem in recent years is owed to the significant progress made on the theoretical and algorithmic side, but also the increase of the computing power of hardware.
In fact, almost all modern quantum chemical techniques rely on a compact representation (i.e., an effective storage) and efficient manipulation of the many-body wave function~\cite{coester1958,coester1960,Bartlett2007b,white1992,ostlund1995,Booth2009a,holmes2016} or the corresponding reduced density matrices~\cite{nakatsuji1976,mazziotti1998,mazziotti2004,piris2007,PG16}.
In particular, for weakly correlated systems, efficient and accurate solutions
are now routinely obtained even at large scales~\cite{rolik2011a,rolik2013,ma2017,kurian2024}. In contrast, the problem of strong correlations remains a critical challenge.
A promising direction to address and solve this problem might be offered by quantum computers as part of the ongoing second quantum revolution~\cite{dowling2003quantum, atzori2019second, deutsch2022second}. However, the extent to which quantum computers can assist in solving the strong correlation problem is still under debate~\cite{CRODJKKMP19, MPJMMHP19, MEABY20,lee2023a}.
To address the strong correlation problem, various heuristic schemes have been designed to compress the wave function~\cite{white1992,limanni2020,liao2024unveiling} or to reduce the original problem to a simpler one, e.g., through embedding techniques~\cite{georges1996a,GC12, KC13, wolf2015a,SC16, SB18,zgid2011,STSF21} and effective Hamiltonian methods~\cite{lowdin1963,white2002,neuscamman2010a,evangelista2014,Ochi2017,Luo2018,Dobrautz2019,liao2021c,liao2023,bauman2023,kowalski2023}.  In the meantime, measures have been developed that quantify the strong
correlations from different perspectives~\cite{tishchenko2008, lee1989,janssen1998,stein2017a} in order to better understand the electron correlation problem and its complexity. Inspired by these works, we believe that a comprehensive understanding of the underlying
correlation structure is imperative to facilitate the development of more efficient schemes for compressing and reducing the complexity of the correlation problem.

\begin{figure*}[htb]
\centering
\frame{\includegraphics[width=1.0\linewidth]{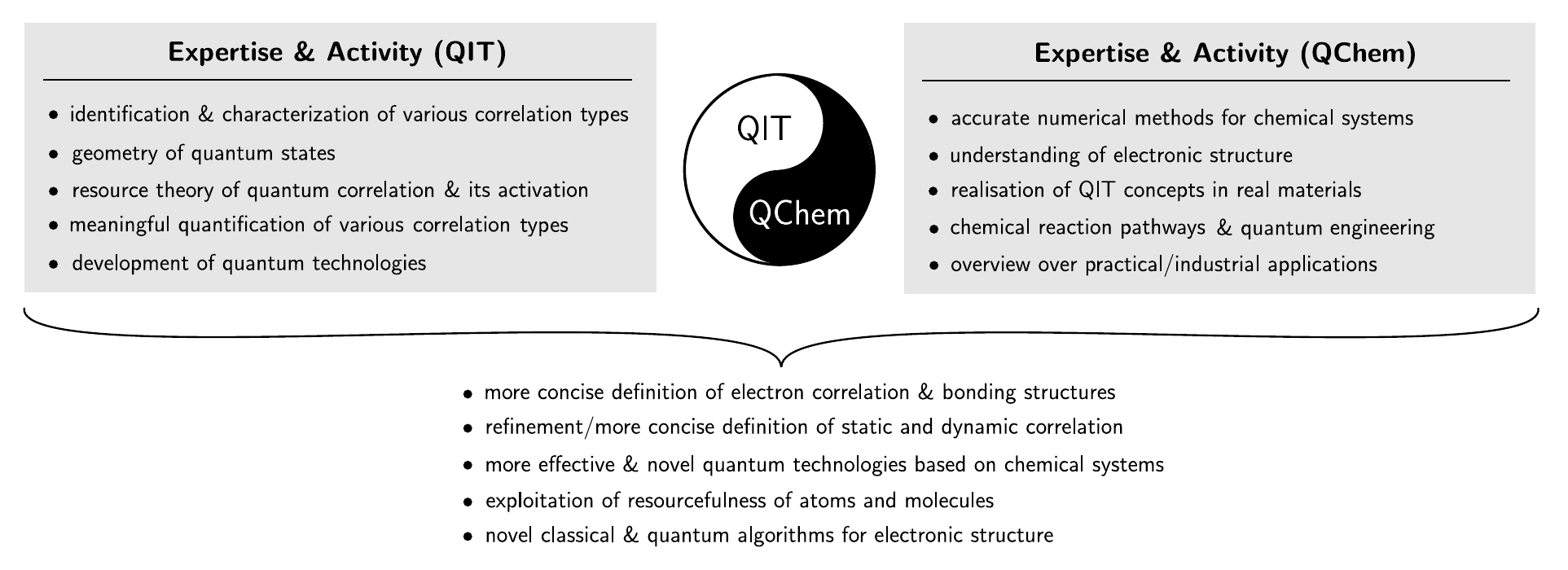}}
\caption{Illustration of the expertise and activities within quantum information theory (QIT) and quantum chemistry (QChem) and anticipated potential  synergy with and emphasis on the electron correlation problem and the ongoing second quantum revolution (see text for more details). \label{fig:summary}}
\end{figure*}

At the same time, the concepts of correlation and complexity lie also at the heart of quantum information theory (QIT). With distinguishable parties as subsystems, concise and operationally meaningful characterizations of various correlation types have long been established, the most famous one being the entanglement~\cite{werner1989quantum,vedral1997quantifying,henderson2001,groisman2005}.
Together with the geometric picture of quantum states~\cite{bengtsson2017geometry}, different correlation types can be elegantly unified under the same theory~\cite{modi2010unified}. Such characterization of correlation is mathematically rigorous, thus offering precise assessment of the fermionic correlation in quantum systems. Crucially, it is also operationally meaningful, in that QIT quantifies correlation and entanglement as the exact amount of available resource in quantum systems for distinctive information processing tasks such as quantum teleportation~\cite{PhysRevLett.70.1895,nielsen2002quantum} and superdense coding~\cite{PhysRevLett.69.2881,nielsen2002quantum}. As we transition into the second quantum revolution, the practical aspects of quantum technologies also emerge as major challenges that cannot be tackled by the field of QIT alone. In particular, physical realization of qubits using molecular systems, storing, and manipulating quantum information therein, are all on-going interdisciplinary endeavors.

We illustrate and summarize the resulting interplay between QIT and QChem based on their different expertise and research activities in Fig.~\ref{fig:summary}. Given the distinct strengths of both fields and their needs and long-term goals, the two fields can complement each other and form a powerful synergy. On one hand, QIT offers precise characterizations of various aspects of electron correlation, which could simplify descriptions of correlated many-electron wave functions, refine our understanding of static and dynamic correlation, and even inspire new (classical or quantum) approaches towards the electron correlation problem~\cite{eisert2016orbital,stein2016automated,ding2023b,liao2024unveiling}. In order to achieve any of these, it is essential to adapt the correlation and entanglement theory from QIT, which was designed specifically for distinguishable systems, to systems of indistinguishable electrons. This is indeed  a nontrivial task, as major theoretical considerations are involved regarding fermionic antisymmetry, superselection rules~\cite{wick1970superselection,SSR} or the $N$-representability problem~\cite{Coleman-book}, amongst others. On the other hand, expertise from the QChem is absolutely essential for the development of effective and novel quantum registers based on atoms and molecules, and for exploiting the resourcefulness of the entanglement therein.

It is therefore of great importance that the two communities join forces and open up a communication line for active discussions, which is exactly the purpose of this work. Our article focuses more on how quantum chemistry can benefit from QIT, and we structure it as follows. In Section \ref{sec:QIparadigm}, we revisit the key concepts of the geometry of quantum states, as well as the entanglement and correlation in systems of distinguishable particles as studied in QIT. In Section \ref{sec:fermions2ndQ} and \ref{sec:fermions1stQ}, we explain how one can adapt these concepts to the setting of indistinguishable fermions, in both the orbital and particle picture, respectively. In Section \ref{sec:examples}, we demonstrate several applications of using fermionic entanglement and correlation as tools for simplifying the structure of molecular ground states.

\section{The Quantum Information Paradigm}\label{sec:QIparadigm}
In this section we introduce some notation and recall basic aspects of quantum information theory. These are the crucial concepts of correlation and entanglement in composite quantum systems and the underlying paradigm of local operations and classical communication, discussed in the context of bipartite systems.

\subsection{Quantum systems and quantum states}\label{subsec:systems}

We start by considering a complex finite-dimensional Hilbert space $\mathcal{H}$ of dimension $d$ and denote the algebra of linear operators acting on $\mathcal{H}$ by $\mathcal{B}(\mathcal{H})$. Quantum states are then described by operators $\rho$ on $\mathcal{H}$ which are Hermitian, positive semidefinite (i.e., all eigenvalues are nonnegative) and trace-normalized to unity. The corresponding set of all density operators,
\be\label{eq:set-density-op}
\DD(\HH) =\{\rho\in \BH\,|\, \rho^\dagger =\rho,\, \rho \geq 0, \, \Tr[\rho]=1\}\,,
\ee
is convex as any convex combination $\rho=p\rho_1+(1-p)\rho_2$, $p\in [0,1]$, of any two states $\rho_1, \rho_2 \in \DD(\HH)$  belongs to $\DD(\HH)$ as well. In the following, provided there is no ambiguity, we will denote this set just by $\DD$. The boundary of $\mathcal{D}$ is given by those $\rho$ which have at least one zero eigenvalue or, equivalently, which are not of full rank. Pure states are by definition the extremal points of the set $\mathcal{D}$, i.e., those elements that cannot be expressed as a proper convex combination of other points. They are precisely those boundary points which are rank-one projectors, $\rho=\ket{\psi}\!\bra{\psi}$. We briefly illustrate all these crucial aspects for the qubit, i.e., a quantum system with Hilbert space $\mathcal{H} \cong \mathbbm{C}^2$. Any qubit quantum state $\rho$ can be parameterized as
\be
\rho=\frac{\mathbbm{1}+\sum_{i=1}^3n_i\sigma_i}{2}
\ee
with $\vec{n}:= (n_1,n_2,n_3)$ satisfying $|\vec{n}|\leq 1$ and $\{\sigma_i\}$ is the set of Pauli matrices. In the three-dimensional $\vec{n}$-space the set $\DD$ of quantum states takes the form of ball centered around $(0,0,0)$ (Bloch ball~\cite{nielsen2002quantum}).
Its center corresponds to the maximally mixed state $\frac{\mathbbm{1}}{2}$ and the boundary of the ball, characterized by $|\vec{n}|=1$, contains only pure states. For systems with a Hilbert space of dimension larger than two, $\DD$ does not take the form of a ball anymore and most boundary points are not pure anymore (see, e.g., textbook\cite{bengtsson2017geometry}).

Equipped with a basic notion of quantum states we can introduce expectation values of observables $O\in \mathcal{B}(\mathcal{H})$ as $\langle O\rangle_{\rho}=\Tr[O\rho]$.
It is instructive to interpret this as a complex linear functional
\begin{equation}\label{eq:w}
\omega:\mathcal{B}(\mathcal{H})\to \mathbbm{C}\,,
\end{equation}
which is positive semidefinite ($\omega(O^\dagger O)\geq0$) and normalized ($\omega(\mathbbm{1})=1$). Density operators $\rho \in \DD$ turn out to be in a one-to-one correspondence to such linear maps according to $\omega_\rho(O)=\Tr[O\rho]$\cite{Landsman2017FoundationsOQ}. 
As discussed below, this more abstract notion of quantum states  in terms of complex linear maps has the advantage that it allows one to define in composite systems the concept of reduced states quite elegantly. In this approach,
the notion of Hilbert spaces emerges only a posteriori from the Gelfand-Naimark-Segal (GNS) construction~\cite{Landsman2017FoundationsOQ,segal1947irreducible,gelfand1943imbedding}. 
Last but not least, the space $\mathcal{B}(\mathcal{H})$ can be equipped with the Hilbert-Schmidt inner product $\langle O_1,O_2\rangle:=\Tr[O_1^\dagger O_2]$,  $O_1,O_2\in \mathcal{B}(\mathcal{H})$. This in turn induces a norm (Frobenius norm) which can then be used to quantify distances between linear operators in general and quantum states in particular, according to $d_F(\rho,\sigma):= ||(\rho-\sigma)||_F=\sqrt{\Tr[(\rho-\sigma)^2]})$.

\subsection{Subsystems and reduced density operators}\label{subsec:subsystems}

The concept of subsystems plays a pivotal role in the quantum sciences in general. For instance, conventional quantum computing exploits quantum effects in a multipartite system comprised of qubits. Also in physics and chemistry it is often necessary to regard the system of interest just as a subsystem of a larger one, e.g., due to its interaction with an environment. Also the electrons and nuclei can be considered as subsystems of an atom, while atoms are subsystems of molecules. We now briefly recall the theory of bipartite systems made up of two subsystems which are distinguishable, such as the paradigmatic system of two distant labs $A$ and $B$, or two particles of different species.

The Hilbert space $\HH_{AB}$ of such bipartite systems is given as the tensor product
\be \label{eq:hilbertspace}
\HH_{AB} = \HH_A \otimes \HH_B
\ee
of the Hilbert spaces $\HH_{A/B}$ of `Alice and Bob'. 
At the level of operators, the compoundness of such systems translates to the relation
\be\label{hilb}
\BHAB = \BHA \otimes \BHB \; .
\ee
We first focus on one of the two subsystems only, say, Alice's one, and discuss in the next section the interplay between both subsystems. Resorting to the abstract and more elegant notion of quantum states \eqref{eq:w}, one identifies the joint system with the algebra $\BHAB$ of operators and Alice's subsystem with its subalgebra
\be \label{eq:alicessubalgebra}
\{ O_A  \otimes \mathbbm{1}_B , O_A \in \BHA \} \subset \BHAB \; .
\ee
Indeed, the latter is closed under taking linear combinations, products and the adjoint. From a more fundamental point of view, it has actually been established by Zanardi~\cite{zanardi} that general subalgebras are precisely those mathematical objects that define subsystems.

Focusing on operators rather than on vectors in the Hilbert space has an immediate advantage when one defines reduced states of subsystems. For our setting, a given state $\omega_{AB} : \BHAB \rightarrow \mathbbm{C}$ can be `reduced' to a state $\omega_A: \BHA \rightarrow \mathbbm{C}$ by restricting the action of $\omega$ to Alice's subalgebra, see Eq.~\eqref{eq:alicessubalgebra}, i.e., to operators taking the form $O_A  \otimes \mathbbm{1}_B$:
\be \label{eq:reducedomega}
\omega_A (O_A) := \omega_{AB} (O_A  \otimes \mathbbm{1}_B) \; .
\ee
As stated in Sec.~\ref{subsec:systems}, one can then univocally associate to $\omega_A$ a density operator $\rho_A \in \BHA$ by requiring that the equality $\omega_A(O_A) = \Tr[\rho_A O_A]$ is valid $\forall \, O_A \in \BHA$. If $\rho_{AB}$ is the density operator corresponding to $\omega_{AB} $, the defining equality \eqref{eq:reducedomega} then corresponds to
\be \label{eq:parttrace}
\Tr_A[\rho_A O_A] = \Tr_{AB}[ \rho_{AB} O_A \otimes \mathbbm{1}_B ] \; .
\ee
The reduced density operator $\rho_A$ turns out to be the familiar partial trace (over the complementary subsystem $B$) of $\rho_{AB}$, i.e., $\rho_A = \Tr_B[\rho_{AB}]$.

\subsection{Independent subsystems and correlations}\label{subsec:independentsubs}

So far we revisited the mathematical framework for studying a single quantum system and introduced a notion of subsystems. If one is interested in the interplay of two or more subsystems a few more tools are needed which we are going to introduce in the following. This will then allow us in the subsequent section to define and quantify in rigorous terms correlation and entanglement between subsystems.

In analogy to Alice subsystem, Bob's subsystem $B$ is associated with the subalgebra of operators $\mathbbm{1}_A \otimes O_B$, where $O_B \in \BHB$. An important remark is that any two `local' operators $O_A \otimes \mathbbm{1}_B$ and $\mathbbm{1}_A \otimes O_B$ commute. This property of commutativity has been identified~\cite{zanardi} as the defining property for a general notion of \emph{independent subsystems}: correlations between subsystems are understood as correlations between the outcomes of joint but \emph{independent} measurements on subsystems $A$ and $B$. The independence corresponds on the mathematical level to precisely the commutativity of the `local' algebras of observables.

\begin{figure*}[htb]
\centering
\includegraphics[width=1.0\linewidth]{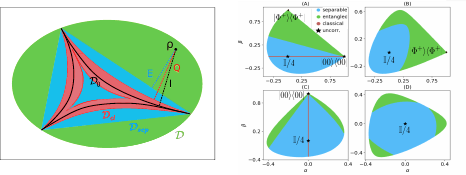}
\caption{Left: schematic illustration of the sets of uncorrelated (black), classical (red), separable (blue) and entangled (green) states and the corresponding geometric correlation measures. Right: two-dimensional intersections of the high-dimensional set of two-qubit states. The point $(\alpha,\beta)=(0,0)$ represents the maximally mixed state. While in (A), (B), (C) the underlying intersecting plane contains the states $\ket{00}$ and/or $\ket{\Phi^+} = (\ket{00}+\ket{11})/\sqrt{2}$, the intersecting plane for (D) is chosen at random.\label{fig:state-manifolds}}
\end{figure*}

While correlations are conceptually rooted in the concept of (local) measurements and operators, they are encoded in terms of properties of the system's quantum states. Specifically, different types of correlations are related to different types of states (uncorrelated states, entangled states, etc), while the amount of correlations can be assessed by correlation measures, i.e., functions $M(\rho)$ of the density operator $\rho$.

\subsection{Hierarchy of states and correlation measures\label{subsec:hierarchy}}

The concise notion of correlation and entanglement is based on the notion of subsystems as discussed in Sec.\ref{subsec:subsystems} and~\ref{subsec:independentsubs}.
As a motivation for the upcoming definition of uncorrelated states, we consider two local operators, $O_A\in \mathcal{B}(\mathcal{H}_A)$ and $O_B\in \mathcal{B}(\mathcal{H}_B)$, and their corresponding correlation function for a given state $\rho_{AB}$,
\be\label{corr}
\begin{split}
\mathcal{C}(O_A,O_B):=\langle O_A\otimes O_B\rangle_{\rho_{AB}}-\langle O_A\rangle_{\rho_A}\langle O_B\rangle_{\rho_B}
\end{split}\,.
\ee
If for some pair of $O_A,O_B$ the above function $\mathcal{C}$ vanishes, it can still be non-zero for some other pair $O'_A,O'_B$. This observation suggests that a state $\rho_{AB}$ is \emph{uncorrelated} if and only if its correlation function vanishes for \emph{any} pair of local observables. The states for which this holds true are exactly those of the form $\rho_{AB}=\rho_A\otimes\rho_B$, the so-called product states, forming the set
\be
\mathcal{D}_0:=\{\rho_{AB}=\rho_A\otimes\rho_B\}~ .
\ee
These states are precisely those that Alice and Bob can prepare through local operations. If Alice and Bob are in addition allowed to communicate classically, i.e., if we refer to the scheme of `local operations and classical communication ' (LOCC) \cite{Chitambar_2014,werner1989quantum}, they can create probabilistic mixtures of such uncorrelated states. These are the so-called `separable' or `unentangled' states which form the set
\be\label{dsep}
\mathcal{D}_{sep}:=\left\{  \rho_{AB} \!=\! \sum_{i} p_{i} \rho_A^{(i)} \otimes \rho_B^{(i)}, p_{i}\! >\! 0,\sum_{i} p_{i}\!=\!1 \right\}.
\ee
In fact, $\mathcal{D}_{sep}$ is the convex hull of the set $\mathcal{D}_0$, with the extremal points given by the uncorrelated pure states $\ket{\psi}_A\bra{\psi}_A\otimes\ket{\phi}_B\!\bra{\phi}_B$. In the following, we skip the subscript AB of $\rho_{AB}$ whenever it is clear from the context that we refer to the joint quantum state.

Based on the definition of uncorrelated and unentangled states one can now introduce measures of correlation and entanglement by quantifying, e.g., through the quantum relative entropy~\cite{vedral1997quantifying,nielsen2002quantum,Lindblad74}
\be
S(\rho||\sigma) :=  \Tr[\rho(\log_2{\rho}-\log_2\sigma)]\,,
\ee
how far away a given state $\rho$ lies from the sets $\mathcal{D}_0$ and $\mathcal{D}_{sep}$, respectively. The quantum relative entropy  quantifies how difficult it is to distinguish $\rho$ from $\sigma$~\cite{vedral2002role,hiai1991proper} and enjoys a number of useful properties such as convexity in both arguments, unitary invariance and it fulfills the `data processing' inequality~\cite{Tomamichel_2016}. This in turn leads to the following measures of correlation $I$ \footnote{Note that $I$ is also sometimes referred to as `total correlation' to highlight that it contains in general both classical and quantum correlations.} and entanglement $E$,
\be\label{entcorr}
\begin{split}
    I(\rho)&:=\min_{\sigma\in\mathcal{D}_0} S(\rho||\sigma)\,,\\
    E(\rho)&:=\min_{\sigma\in\mathcal{D}_{sep}} S(\rho||\sigma)\,.\\
\end{split}
\ee
The geometrical nature of the measures in \eqref{entcorr} as well as the different state manifolds are graphically illustrated in Fig.~\ref{fig:state-manifolds}.

One remarkable fact is that the minimization underlying $I$ in \eqref{entcorr} can be performed explicitly. Given a state $\rho_{AB}$ the minimizer is the product state $\rho_A\otimes\rho_B$~\cite{modi2010unified}. This yields
\begin{eqnarray}\label{correxp}
I(\rho_{AB}) &=&S(\rho_{AB}||\rho_A\otimes\rho_B) \nonumber \\
&=&S(\rho_A)+S(\rho_B)-S(\rho_{AB})\,,
\end{eqnarray}
where $S(\rho)=-\Tr[\rho\log_2{\rho}]$ is the von Neumann entropy.

Another important aspect of the definition of $I$ is that it universally bounds from above~\cite{wolf2008area,Watrous,CS21lecture} the correlation function of Eq. \eqref{corr} according to
\begin{equation}\label{CorrFversusI}
\frac{|\mathcal{C}(O_A,O_B)|}{||O_A||_F ||O_B||_F} \leq  \sqrt{2}\sqrt{I(\rho_{AB})}\,.
\end{equation}
The bound \eqref{CorrFversusI} quantitatively confirms our intuition: whenever a state is close to $\mathcal{D}_0$ then for any choice of local observables $O_A,O_B$ the correlation function is small. In particular, when the correlation $I$ is zero, then for all possible local operators the correlation function vanishes. A large value of correlation $I$ is generally regarded unfavorable from a computation viewpoint, as such a state would require a larger amount of computational resources for preparing, storing and manipulating it. To the contrary, from the viewpoint of quantum information, such states are resourceful (see Sec.~\ref{subsec:resource}) and, thus, can be used to realize quantum information processing tasks, e.g., in quantum communication or quantum cryptography \cite{Ekert91, BCMdW10, Chitambar_2019}. Unlike for $I$, there does not exist a closed form for the (quantum relative entropy of) entanglement $E$ \eqref{entcorr}.
Nonetheless, it can be calculated for certain states which possess a large number of symmetries~\cite{Vollbrecht_2001}. In particular, for pure states $\ket{\Psi}$ on $\mathcal{H}_{AB}$ there exists a well-known closed form\cite{vedral1997quantifying}
\be \label{eq:entpurestates}
E(\ket{\psi}\!\bra{\psi})=S(\rho_A)=S(\rho_B)=\frac{I(\ket{\psi}\!\bra{\psi})}{2}.
\ee
For general states, one of the crucial properties of $E$ is that it does not increase under LOCC~\cite{plenio2006introduction} operations. This means when the parties $A,B$ are restricted to LOCC, they can only degrade the entanglement content of their state. This relates nicely to the idea that entanglement is a resource which is useful for quantum information processing tasks.

Entanglement is not the only form of correlation that exists. Separable states can possess yet another type of correlation which is useful in quantum information protocols. Such correlations are called quantum correlations beyond entanglement. To be more specific, we first define the set of \emph{classical states},
\be\label{dcl}
\mathcal{D}_{cl}:=\Big\{\,\rho\!=\!\sum_{i,j}p_{ij} |i\rangle\!\langle i|\!\otimes\! |j\rangle\!\langle j| \,\Big\},
\ee
where $\{\ket{i}\}$ and $\{\ket{j}\}$ are any two orthonormal bases of $\mathcal{H}_A$ and $\mathcal{H}_B$ respectively and $p_{ij}$ is some joint probability distribution, with $p_{ij}\geq 0$ and $\sum_{ij}p_{ij}=1$. In other words, a state $\chi$ is classical, $\chi\in\mathcal{D}_{cl}$, if and only if it is diagonal in some product basis set $\{\ket{i}\otimes\ket{j}\}$ for $\mathcal{H}_A\otimes\mathcal{H}_B$. These states are purely classical in the sense that their correlation structures can be understood on a purely classical level~\cite{oppenheim2002thermal,luo2008classical,modi2010unified}.

Given the set $\mathcal{D}_{cl}$, we again use the quantum relative entropy to quantify the `distance' of a given state $\rho$ to that set. This leads to the following definition of quantum correlation:
\be\label{qcorr}
Q(\rho):=\min_{\sigma\in \mathcal{D}_{cl}}S(\rho||\sigma).
\ee
Similar to $E(\rho)$, there is no closed formula for $Q(\rho)$ in general. 
The quantity $Q$ is sometimes regarded as symmetric discord or geometric discord and plays an important role for the realization of tasks such as quantum state merging~\cite{PhysRevA.83.032323} or quantum key generation~\cite{WZ2015}.
Moreover, quantum correlation can be converted to entanglement, $E$, via distinct activation protocols~\cite{PhysRevLett.106.220403}.

Note that any state of the form $\rho_A\otimes\rho_B$ can be written as $(\sum_i p^A_i\ket{i}\!\bra{i})\otimes (\sum_j p^B_j\ket{j}\!\bra{j})=\sum_{ij}p^A_ip^B_j\ket{ij}\!\bra{ij}$, which is of the form~\eqref{dcl}. Furthermore, it is clear by definition \eqref{dcl} that classical states are in particular separable. This leads to the following inclusion hierarchy
\be
\mathcal{D}_0\subset\mathcal{D}_{cl}\subset\mathcal{D}_{sep}\,,
\ee
which together with \eqref{entcorr}, \eqref{qcorr} directly implies
\be
I(\rho)\geq Q(\rho)\geq E(\rho) \; .
\ee
The hierarchy of the various sets of states and the geometric notion of the measures is presented in Fig. \ref{fig:state-manifolds}. Given the definition of quantum correlation contained in a state $\rho$ one can also quantify the amount of classical correlation contained in a state $\rho$. For this one first finds the closest classical state $\chi_\rho\in\mathcal{D}_{cl}$ which fulfills $Q(\rho)=S(\rho||\chi_\rho)$. Then the amount of classical correlation in $\rho$ is defined as $C(\rho)=I(\chi_\rho)$, i.e., the correlation contained in the state after all quantum correlation has been extracted \cite{modi2010unified}.

The above geometric ideas straightforwardly extend~\cite{modi2010unified} to systems composed of $N>2$ distinguishable subsystems. Here the set of uncorrelated states contains density operators of form $\rho=\bigotimes_{i=1}^N\rho_i$. The fully separable states are given by convex combination of uncorrelated states and classical states are given by $\rho=\sum_{\vec{k}}p_{\vec{k}}\ket{\vec{k}}\bra{\vec{k}}$, where $\{\ket{\vec{k}}:=\bigotimes_{i=1}^N\ket{k_i}\}$ is any product basis for $\bigotimes_{i=1}^N\mathcal{H}_i$. As a generalization of Eq.~\eqref{correxp} the corresponding correlation $I$ can be explicitly evaluated and follows as
\be\label{totcordsub}
I(\rho)=\sum_{i=1}^N S(\rho_i)-S(\rho)\,,
\ee
where $\rho_i$ is the reduced state of subsystem $i$ \cite{modi2010unified}.

\subsection{Entanglement as a resource}\label{subsec:resource}

To highlight in particular that entanglement plays the role of a key resource for quantum information processing tasks, we discuss two important protocols that form the building blocks of several other protocols in quantum information theory. The first protocol is that of quantum teleportation~\cite{PhysRevLett.70.1895,nielsen2002quantum}, which enables Alice to transmit an \textit{unknown} quantum state $\ket{\psi}_a$ to Bob. It relies on the use of entanglement and LOCC. To explain this, suppose Alice and Bob are spatially separated and share a maximally entangled state of two electrons (or two qubits), given as $\ket{\phi^+}_{AB}=(\ket{00}+\ket{11})/\sqrt{2}$ where $\ket{0}$ refers to spin up and $\ket{1}$ refers to spin down. In addition Alice possesses another electron in an unknown state $\ket{\psi}_{a}=\alpha\ket{0}+\beta\ket{1}$. Accordingly, the joint state of the three electrons can be written as
\be\label{tele}
\begin{split}
\ket{\psi}_a\otimes\ket{\phi^+}_{AB}=&\frac{1}{2}\left[\ket{\phi^+}_{aA}\otimes\ket{\psi}_B+\ket{\phi^-}_{aA}\otimes(\sigma_3\ket{\psi}_B)\right.\\
&+\ket{\psi^+}_{aA}\otimes(\sigma_1\ket{\psi}_B)\\
&\left.+\ket{\psi^-}_{aA}\otimes(\sigma_3\sigma_1\ket{\psi}_B)\right]\,.
\end{split}
\ee
where we introduced the orthonormal basis $\mathbf{BELL}:=\{\ket{\phi^{\pm}}_{aA}=(\ket{00}\pm\ket{11})/\sqrt{2}, \; \ket{\psi^\pm}_{aA}=(\ket{01}\pm\ket{10})/\sqrt{2} \}$ for $\mathbbm{C}^2\otimes\mathbbm{C}^2$. 
From \eqref{tele} it is apparent that the outcome of Alice's local measurement on $a,A$ in the Bell basis identifies a distinctive unitary that Bob afterwards could apply in order to get the unknown state $\ket{\psi}_B$ of Alice. For example, if Alice measures $\ket{\phi^-}_{aA}$ then Bob will need to apply the Pauli matrix $\sigma_3$ on his system.
This means that Bob is able to aptly recover Alice's unknown state after Alice has communicated classically her measurement outcome to him through two bits of information.

To discuss a second important protocol we consider Alice and Bob being connected via a quantum communication channel. Alice holds a single electron that she can send to Bob in order to transmit information. Without additional resources, the best she can do is to encode the classical bit (0 or 1) onto the spin of the electron and pass it to Bob via the channel who then measures the spin along the $z$-axis to determine the value of the classical bit. In this way Alice can send one classical bit of information to Bob. However, if Alice and Bob share in addition an entangled state $\ket{\phi^+}_{AB}$ then Alice can communicate two bits of information. To see this notice that the set of states $\{\sigma^A_i\otimes\mathbbm{1}_B\ket{\phi^+}_{AB}\}_{i=0}^3$ coincides with $\mathbf{BELL}$ up to phase factors and thus forms an orthonormal basis for $\mathbbm{C}^2\otimes\mathbbm{C}^2$ (here $\sigma_0=\mathbbm{1}$). This means Alice can choose to apply one of the four unitary operators $\sigma_i$ to her system and send her electron afterwards to Bob. Bob then performs a joint measurement on both electrons in the Bell basis to determine which of the four operators was applied by Alice. Accordingly, Alice could encode two classical bits using the four $\sigma_i$ operators and Bob can decode by performing a Bell measurement. This protocol is called superdense coding~\cite{PhysRevLett.69.2881,nielsen2002quantum}.

If we denote a unit  of entanglement ($\ket{\phi^+}$) by $[qq]$, a quantum channel that transmits single qubits by $[q\to q]$ and a classical channel that transmits one bit by $[c\to c]$ then the above protocols can concisely be summarized as:
\be
\begin{split}
    [qq]+2[c\to c]&\geq [q\to q] \mbox{ (quantum teleportation)}\\
    [qq]+[q\to q]&\geq 2[c\to c] \mbox{ (superdense coding)}\,.
\end{split}
\ee
In these so-called resource inequalities~\cite{Devetak_2008} the sign $\geq $ emphasizes that the left hand side is at least as resourceful as the right hand side. To conclude, these two remarkable quantum protocols univocally demonstrate the necessity for a concise and  operationally meaningful quantification of entanglement and various other correlation types.

\section{The fermionic orbital picture}\label{sec:fermions2ndQ}

After having explained some basic concepts and tools from quantum information theory, we adapt them now to systems of identical fermions, and in particular electrons. In Section \ref{subsec:OrbCandE}, we present the fermionic `orbital' picture which is based on the formalism of second quantization and accordingly regards atomic or molecular orbitals as subsystems. Since orbitals are distinguishable (as opposed to the fermions themselves), quantum information theoretical concepts can be applied in a straightforward manner, although the fermionic superselection rule requires some additional care (see Section \ref{subsec:SSR}). The `particle' picture which is based on 1st quantization will be the subject of Sec.~\ref{sec:fermions1stQ}. It is rather delicate since the antisymmetrization removes the mathematical feature that quantum information theory relies on, namely the tensor product structure of Hilbert spaces describing multipartite systems.

\subsection{Notation and formalism}\label{subsec:formalism2ndQ}

We consider a finite dimensional one-particle Hilbert space $\Ho \cong \mathbbm{C}^d$, which is spanned by the elements $\ket{1}, \ket{2},\ldots,\ket{d}$ of a reference orthonormal basis, also called `modes'. One can think of them as a basis of spin-orbitals given by $\ket{\phi_i,\sigma}$ (which we sometimes abbreviate as $\ket{j}$), where $\ket{\phi_i}$ represents a spatial orbital and $\sigma \in \{ \uparrow, \downarrow \}$ is the spin variable. The corresponding Hilbert space for $N \leq d$ fermions is given as the antisymmetrized $N$-fold tensor product of $\Ho$
\be \label{eq:Nfermionspace}
    \HN := \mathcal{A}( \, \underbrace{\Ho  \otimes \ldots \otimes \Ho }_{N \text{ times}} \, )\,.
\ee
We also introduce the Fock space
\be
\F := \CC\oplus \bigoplus_{N=1}^d \HN \; ,
\ee
on which creation (annihilation) operators $f_i^{\dagger}$ ($f_i^{\phantom{\dagger}}$) act by creating (annihilating) a fermion in mode $\ket{i}$. The antisymmetrization involved in Eq.~\eqref{eq:Nfermionspace} translates to the canonical anticommutation relations,
\begin{equation} \label{eq:car}
    \{f^{\phantom{\dagger}}_i,f^{\dagger}_j\} = \delta_{ij}\,, \quad\{f^{\phantom{\dagger}}_i,f^{\phantom{\dagger}}_j\} =\{f_i^{\dagger},f^{\dagger}_j\} = 0 \,,
\end{equation}
where $\{f^{\phantom{\dagger}}_i,f^{\dagger}_j\} = f^{\phantom{\dagger}}_i f^{\dagger}_j + f^{\dagger}_j f^{\phantom{\dagger}}_i$ is the anticommutator.

With the given reference basis $\{ \ket{i}\}$ we can build an `occupational number' basis for the Fock space $\F$, composed of $2^d$ Slater determinant state vectors
\be
\ket{\vec{n}} = \big(f^{\dagger}_1\big)^{n_1} \ldots \big(f^{\dagger}_d\big)^{n_d} \ket{0}= \ket{n_1,\ldots,n_d}\,,
\ee
indexed by a configuration vector $\vec{n} \in \{0,1 \}^d$. Here, $\ket{0}$ denotes the vacuum state. The occupation number basis indicates that there exists more than one potential tensor product structure within $\F$: Any partition of the reference basis into subsets induces a corresponding notion of subsystems. For instance, the partition into $\{\ket{1},\ldots,\ket{d'}\}$ and $\{\ket{d' + 1},\ldots,\ket{d}\}$, induces a respective unitary mapping from $\F$ to the tensor product of the Fock space built on the first $d'$ modes and the Fock space built on the remaining modes:
\be \label{eq:fockorbitalmapping}
\ket{\vec{n}} \mapsto \ket{n_1,\ldots,n_{d'}} \otimes \ket{n_{d' + 1},\ldots,n_d} \; .
\ee
We remark that this mapping depends on the underlying reference basis of $\Ho$, or, at a more abstract level, on the choice of a subspace of $\Ho$. Also, notice that some sign ambiguities in Eq.~\eqref{eq:fockorbitalmapping} must be understood and resolved by means of the parity superselection rule (as explained in Sec.~\ref{subsec:SSR}).

\subsection{Orbital correlation and entanglement}\label{subsec:OrbCandE}

Based on mappings such as that of Eq.~\eqref{eq:fockorbitalmapping}, one can adapt the quantum information theoretical formalism discussed in Sec.~\ref{sec:QIparadigm} to the orbital picture of fermionic systems.

We now discuss in more detail some instances of this formalism in the case of electron systems, where the spin degree of freedom comes into play as a factor $\Hs \cong \mathbbm{C}^2$ in the one-particle space,
\be
\Ho = \Hl \otimes \Hs \; .
\ee
Here the orbital one-particle space $\Hl$ is spanned by a system-specific selection of $d/2$ spatial orbitals $\ket{\phi_i}$. The corresponding reference basis of $\Ho$ (see Sec.~\ref{subsec:formalism2ndQ}) is made up of the spin-orbitals $\ket{\phi_i, \sigma}$. We now look at mappings of the form of Eq.~\eqref{eq:fockorbitalmapping}. These are useful for describing, e.g., a lattice of atoms which hosts electrons as a union of two half-lattices regarded as subsystems, a diatomic molecule as a system of two atoms, the cloud of electrons around a nucleus as a union of a set of inner orbitals and a set of outer ones. One then often thinks of a partition of the $d/2$ orbitals spanning $\Hl$ at first, e.g. into three subsets $\{\ket{\phi_1},\ldots,\ket{\phi_{i_1}}\}$, $\{\ket{\phi_{i_1+1}},\ldots,\ket{\phi_{i_2}}\}$ and $\{\ket{\phi_{i_2+1}},\ldots,\ket{\phi_{d/2}}\}$. This in turn induces a corresponding partition of the $d$ spin orbitals spanning $\Ho$.

Particular relevant partitions of the reference basis of the $d$ spin-orbitals $\ket{\phi_i,\sigma}$ are the following ones:
\begin{itemize}
    \item finest partition: each of the $d$ spin-orbitals $\ket{\phi_i,\sigma}$ constitutes a subsystem
    \item finest orbital partition: for $i=1,\ldots,d/2$ each pair $\{\ket{\phi_i,\uparrow},\ket{\phi_i,\downarrow}\}$ constitutes a subsystem, namely the one of the $i$-th spatial orbital
    \item 1 vs.~rest: one (spin-)orbital $\ket{\phi_i}$ ($\ket{\phi_i,\sigma}$) defines a two-mode (one-mode) subsystem; all other \mbox{(spin-)}orbitals form the second subsystem
    \item 1 vs.~1 vs.~rest: two (spin-)orbitals are identified as two subsystems; all other (spin-)orbitals form the third subsystem
    \item closed (doubly occupied) vs. active vs. virtual (empty) orbitals: this general tripartition underlies the idea of complete active spaces.
\end{itemize}
In particular, for valence bond theory the partition of choice is the one where two (orthonormalized) atomic orbitals $i$ and $j$ are singled out as subsystems, and all other orbitals constitute a third subsystem to be discarded~\cite{toolbox}. For $\ket{\vec{n}} = \ket{n_{1\uparrow},n_{1\downarrow},\ldots,n_{d/2,\uparrow},n_{d/2,\downarrow}}$, Eq.~\eqref{eq:fockorbitalmapping} then is adapted according to
\be \label{eq:mappingtwoorbitals}
\ket{\vec{n}}\mapsto \ket{n_{i\uparrow},n_{i\downarrow}} \otimes \ket{n_{j\uparrow},n_{j\downarrow}} \otimes \ket{\text{rest}}\,.
\ee
This mapping associates to each of the two orbitals of interest a `small' Fock space spanned by just four state vectors characterized by spin occupancies: $\ket{0}_{i/j}$, $\ket{\!\uparrow}_{i/j}$, $\ket{\!\downarrow}_{i/j}$, and $\ket{\!\uparrow \downarrow}_{i/j}$. The idea of discarding orbitals $\ket{\phi_k}$ with $k \neq i,j$ corresponds to a partial trace over all such orbitals, whose result is the reduced state $\rho_{ij}$ of the two orbitals.

The two-orbital reduced state $\rho_{ij}$ describes a bipartite system made up of two distinguishable subsystems, namely orbital $\ket{\phi_i}$ and orbital $\ket{\phi_j}$. By evaluating the correlation measures introduced in Sec.~\ref{subsec:hierarchy} on $\rho_{ij}$, one can quantify the correlations between the two orbitals. Such correlations in turn describe how bonded the orbitals $\ket{\phi_i}$ and $\ket{\phi_j}$ are. As an example, we consider a pure state $\rho_{ij} = \ket{\Psi}\!\bra{\Psi}$, where $\ket{\phi_i}$ and $\ket{\phi_j}$ host two electrons with opposite spin in the `bonding' orbital $\frac{1}{\sqrt{2}}(\ket{\phi_i} + \ket{\phi_j})$. Then, using Eq.~\eqref{eq:mappingtwoorbitals}, we can identify $\ket{\Psi}$ with an element of the tensor product of the two Fock spaces of orbitals $\ket{\phi_i}$ and $\ket{\phi_j}$ mentioned above. We obtain
\begin{align} \nonumber
\ket{\Psi} &= \frac{1}{\sqrt{2}} (f^{\dagger}_{i\uparrow} + f^{\dagger}_{j\uparrow}) \frac{1}{\sqrt{2}} (f^{\dagger}_{i\downarrow} + f^{\dagger}_{j\downarrow}) \ket{\vec{0}} \\ \nonumber
&= \frac{1}{2} (f^{\dagger}_{i\uparrow} f^{\dagger}_{i\downarrow} +f^{\dagger}_{i\uparrow} f^{\dagger}_{j\downarrow} + f^{\dagger}_{j\uparrow} f^{\dagger}_{i\downarrow} + f^{\dagger}_{j\uparrow} f^{\dagger}_{j\downarrow}) \ket{\vec{0}}\\ \nonumber
&\mapsto \frac{1}{2} (\ket{\!\uparrow \downarrow}_i\otimes \ket{0}_j + \ket{\!\uparrow}_i\otimes \ket{\!\downarrow}_j - \ket{\!\downarrow}_i\otimes \ket{\!\uparrow}_j \\\nonumber
&\quad\,\, + \ket{0}_i\otimes \ket{\!\uparrow \downarrow}_j) \,.\label{eq:bonding_state}
\end{align}
From a quantum information theoretical perspective, one notices that in this representation $\ket{\Psi}$ looks like a maximally entangled state. Using Eq.~\eqref{eq:entpurestates} and ignoring for the moment the role of superselection rules, the entanglement between the two orbitals evaluates to $E(\ket{\Psi}) = \log_2(4) = 2$~\cite{toolbox}.

\subsection{Superselection rules}\label{subsec:SSR}
The discussions leading to Eq.~\eqref{eq:fockorbitalmapping} in Sec.~\ref{subsec:formalism2ndQ} can be abstracted as follows. A bipartition of a reference basis of the one-particle space $\Ho$ into two sets of modes $A$ and $B$ induces a splitting
\be
\Ho = \Ho_A \oplus \Ho_B
\ee
of $\Ho$ into two orthogonal subspaces. This splitting induces a tensor product structure in the Fock space,
\be
\F[\Ho] \cong \F[\Ho_A] \otimes \F[\Ho_B] ~ .
\ee
However, for $A$ and $B$ to qualify as valid subsystems, operators pertaining to modes in $A$ should commute with those pertaining to modes in $B$, as per our discussion in Sec.~\ref{subsec:independentsubs}. This is clearly not the case given that the fermionic creation and annihilation operators associated with the modes in $A$ anticommute with those relative to modes in $B$ (see Eq.~\eqref{eq:car}). This issue is overcome by imposing the so-called fermionic parity superselection rule (P-SSR)~\cite{wick1970superselection,SSR}, a fundamental rule of nature whose violation would actually make superluminal signalling (i.e., communication faster than the speed of light) possible~\cite{johansson2016comment,ding2020concept}. On the level of states, P-SSR `forbids' coherent superpositions of even and odd fermion-number states. On the level of local operators~\cite{wise04fluffy,bartlett2003,schuch2004quantum} it dictates that physical local observables on modes in $A/B$ must always commute with the local parity operator $\mathcal{P}^{A/B}=\Pi^{(A/B)}_{\text{even}}-\Pi^{(A/B)}_{\text{odd}}$, where $\Pi^{(A)}_\tau$ is the projector onto the subspace of local particle number $\tau\in\{\text{even, odd}\}$. Imposing such a commutation rule between local observables and local parity operator selects out observables on $A$ commuting with those on $B$. This leads to a proper description of $A$ and $B$ as subsystems. Operationally, such a rule has a drastic effect on the accessible entanglement and correlation contained in a state $\rho_{AB}$ as now the physically relevant state is given by the superselected version~\cite{wise04fluffy,bartlett2003} $\rho^\text{P}_{AB}$
\begin{equation}
    \rho_{AB}^\text{P} = \sum_{\tau, \tau' = \text{even, odd}} \Pi_\tau^{(A)} \otimes \Pi^{(B)}_{\tau'} \rho_{AB} \Pi^{(A)}_\tau \otimes \Pi^{(B)}_{\tau'} ~ .
\end{equation}
This implies that the various measures of correlations $M=I,E,Q$ discussed in Sec.~\ref{subsec:hierarchy} must be replaced by superselected versions of them denoted by $M^{\text{P}}$ where
\be
M^{\text{P}}(\rho_{AB}):=M(\rho_{AB}^\text{P})\,
\ee
and $M^{\text{P}}(\rho_{AB}) \leq M(\rho_{AB})$~\cite{LexinGesa}.

\section{The fermionic particle picture}\label{sec:fermions1stQ}

In Sec.~\ref{sec:fermions2ndQ} the fermionic orbital picture has been introduced. We now turn our attention to the fermionic particle picture. It attempts to identify fermions themselves as subsystems and is therefore based on the formalism of first quantization. Yet, the antisymmetrization of state vectors prevents a straightforward application of the concepts developed in Sec.~\ref{sec:QIparadigm} and instead requires an adaption thereof.

\subsection{Fermions as subsystems?}\label{subsec:fermisubsystem}

In Sec.~\ref{subsec:subsystems} it was mentioned that on an abstract level subsystems are described by subalgebras of the operator algebra of the system in consideration. For a system of $N$ fermions, we consider the algebra $\B(\HN)$ of operators on the $N$-fermion space. One may then wonder whether a single fermion constitutes a conventional subsystem. The underlying question can be made more precise: is there a subalgebra of $\B(\HN)$ which describes a single fermion? It turns out~\cite{nosinglefermionsubalgebra} that the answer to this critical question is unfortunately `no'. A particular promising candidate would have been the subset of `one-particle' observables, i.e., operators of the form
\be \label{eq:1Poperator}
h = \sum_{ij} h_{ij} f^{\dagger}_i f^{\phantom{\dagger}}_j \,.
\ee
This subset, however, is not closed under multiplication. For instance, the product of $f^{\dagger}_i f^{\phantom{\dagger}}_i $ and $ f^{\dagger}_j f^{\phantom{\dagger}}_j  $ is the two-particle operator $f^{\dagger}_i  f^{\phantom{\dagger}}_i f^{\dagger}_j f^{\phantom{\dagger}}_j = -f^{\dagger}_i f^{\dagger}_j f^{\phantom{\dagger}}_i  f^{\phantom{\dagger}}_j$. Therefore, this subset does not qualify as a proper physical subsystem.

The embedding of the $N$-fermion antisymmetric space $\HN$ into the larger space $\otimes^N \Ho$ seemingly allows one to recover a tensor product structure. Yet, this approach is dubious and could easily lead to incorrect conclusions. This can be illustrated by considering two (spin-polarized) electrons which have never interacted and which occupy two orbitals $\ket{\phi_1}$, $\ket{\phi_2}$ localized in far away regions. In fact, the corresponding Slater determinant state of the two electrons can be written as
\be \label{eq:sd_faraway}
\frac{1}{\sqrt{2}}(\ket{\phi_1, \uparrow} \otimes \ket{\phi_2, \uparrow} - \ket{\phi_2, \uparrow} \otimes \ket{\phi_1, \uparrow}) \; .
\ee
At first sight, this looks like an entangled state, in striking contrast to the fact that the two electrons have never interacted. Yet, this is merely an artefact of the misleading embedding into the Hilbert space of $N=2$ \emph{distinguishable} particles: the underlying algebra of observables is still the one of fermionic particles rather than the one of distinguishable particles. Therefore these `exchange correlations' are purely mathematical and do not exist in reality.

$\text{   }$
\begin{figure*}[ht!]
\centering
\centering
\frame{\includegraphics[width=1.0\linewidth]{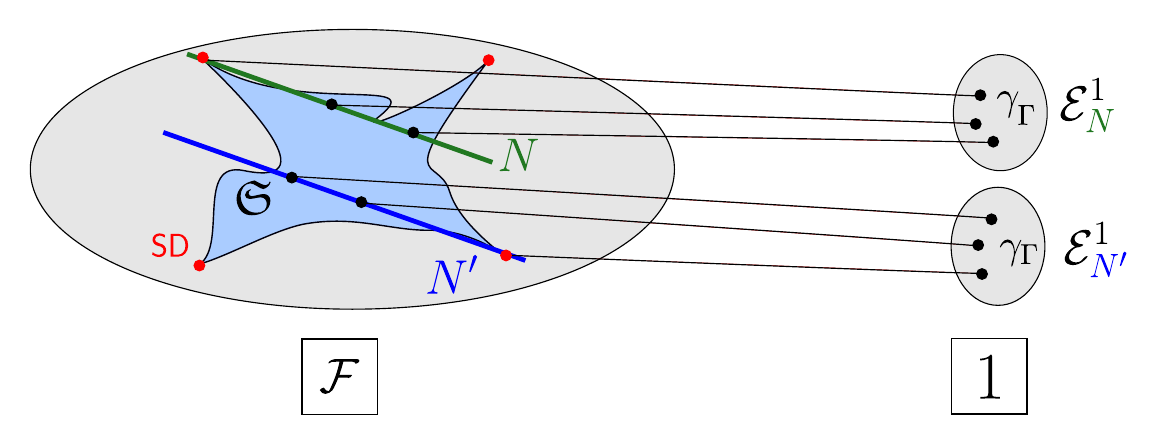}}
\caption{Illustration of the one-to-one relation between the set $\mathfrak{S}$ (light blue) of free states on the Fock space $\F$ (gray) and the sets of 1RDMs $\mathcal{E}^1_N$ for two different particle numbers $N, N^\prime$. The red dots illustrate its extremal elements which are the Slater determinants. For fixed particle number $N$, the free states $\Gamma$ corresponding to 1RDMs $\gamma_\Gamma\in \mathcal{E}^1_N$ lie on a hyperplane with average particle number $\langle \hat{N}\rangle_{\Gamma} = N$.\label{fig:freestates-1RDM}}
\end{figure*}
\subsection{Non-interacting states and nonfreeness\label{sec:free-states}}

Since individual fermions do not qualify as subsystems there does not exist a notion of correlation or entanglement between fermions in a strict quantum information theoretical sense.
However, from the standpoint of quantum chemistry one would like to address correlations as a consequence of interactions between particles. This suggests that if particles (electrons) are subjected to a non-interacting Hamiltonian $h$, see Eq.~\eqref{eq:1Poperator}, the corresponding eigenstates must be deemed as uncorrelated. These eigenstates are just Slater determinants.
If one also considers mixed states, then there is actually a larger class of states which must be regarded as uncorrelated, namely the so-called free states~\cite{A70, Ripka1986, BLS94, gottlieb2007, Gottlieb14}. While Slater determinants are ground states of non-interacting Hamiltonians, free states are the thermal states of such Hamiltonians. The set of free states $\Gamma$ on the Fock space is therefore defined as
\be\label{ferdo}
\mathfrak{S}:=\overline{\left\{\Gamma=\frac{1}{Z}e^{- h}\right\}}\,,
\ee
where $h$ denotes arbitrary one-particle Hamiltonians and the `closure' is required to include Slater determinants as well.
For a `diagonalized' one-particle Hamiltonian $h=\sum_i \varepsilon_i f_i^\dagger f^{\phantom{\dagger}}_i$
the corresponding free state can be represented in the Fock basis $\{\ket{\vec{n}}\}$
as
\be\label{freest}
\Gamma=\frac{1}{Z}\sum_{\vec{n}}e^{-\sum_i n_i \varepsilon_i}\ket{\vec{n}}\bra{\vec{n}}\,.
\ee
The outer summation in \eqref{freest} runs over all possible occupation vectors $\vec{n}$ and the `partition function' ensuring normalization reads $Z=\sum_{\vec{n}}e^{-\sum_i n_i \varepsilon_i}$.
States of the form given by Eq.~\eqref{freest} are also sometimes referred to as number conserving fermionic Gaussian states~\cite{Surace_2022}.

A fundamental property of free states, which in mathematical physics often serves as their actual definition, is that they satisfy a so-called generalized Wick theorem:
every correlation function of a free states splits into a product of two-point correlation functions, which only involves the one particle reduced density matrix (1RDM)~\cite{Robinson65-metric, A70, BLS94, Bach22}.

The 1RDM  of any state $\rho$ on the Fock space follows via
\begin{equation}\label{eq:1RDM-defFock}
\mu^{(1)}: \rho \mapsto \gamma\,,\quad \mbox{with}\,(\gamma_\rho)_{ij}=\Tr[\rho f_j^\dagger f_i^{\phantom{\dagger}}]\,,
\end{equation}
where its matrix elements $(\gamma_\rho)_{ij}$ are calculated in some orthonormal one-particle reference basis. If we restrict $\rho$ to states with fixed particle number $N$, Eq.~\eqref{eq:1RDM-defFock} means nothing else than to trace out $N-1$ particle,
\be\label{eq:1RDM-def}
\gamma_\rho=N\Tr_{N-1}[\rho]\quad \forall\,\rho\in\mathcal{E}^N\,.
\ee
Here and in the following, we denote the set of all ensemble $N$-fermion density operators on $\mathcal{H}^{(N)}$ more conveniently by $\mathcal{E}^N$. The set of 1RDMs that are contained in the image of the set $\mathcal{E}^N$ under the partial trace map in \eqref{eq:1RDM-def} are called ensemble $N$-representable and form the convex set
\begin{equation}\label{eq:E1N}
    \mathcal{E}^1_N :=N\mathrm{Tr}_{N-1}[\mathcal{E}^N]\,.
\end{equation}

In general, 1RDMs do not have a unique preimage in the set $\mathcal{E}^N$ and thus also not in the set of all states on the Fock space $\F$.
This changes considerably, if we restrict ourselves to the free states since they are in a one-to-one correspondence with 1RDMs~\cite{BLS94, Gottlieb14}: for a free state $\Gamma$ \eqref{freest}, the corresponding 1RDM reads
\be
\gamma_\Gamma=\sum_i \lambda_i\ket{i}\!\bra{i}
\ee
where $\lambda_i=\Tr[\Gamma f^\dagger_i f^{\phantom{\dagger}}_i] = 1/(1+\mathrm{exp}(\varepsilon_i))$ are the eigenvalues of the 1RDM. Conversely, for a 1RDM $\gamma=\sum_i \lambda_i\ket{i}\bra{i}$ its corresponding free state follows as (see, e.g., \cite{Gottlieb14})
\be \label{eq:1RDMtofree}
\Gamma=\sum_{\vec{n}}\left\{\prod_i \lambda_i^{n_i}(1-\lambda_i)^{1-n_i} \right\}\ket{\vec{n}}\bra{\vec{n}}\,.
\ee
We illustrate this one-to-one relation between free states and their 1RDMs in Fig.~\ref{fig:freestates-1RDM}. For different particle numbers $N, N^\prime$, the free states mapping to 1RDMs in $\mathcal{E}^1_N$ or $\mathcal{E}^1_{N^\prime}$ lie on hyperplanes of fixed average particle number $\langle \hat{N}\rangle_{\Gamma} = N, N^\prime$.

Equipped with the definition of free states, 
one defines the so-called nonfreeness ~\cite{Gottlieb_2005, Gottlieb14}
\be\label{nonfreeness}
\mathcal{N}(\rho):=\min_{\sigma\in\mathfrak{S}} S(\rho||\sigma)
\ee
for a state $\rho$ on $\F$. Thus, the nonfreeness $\mathcal{N}$ measures how far a state is from the set $\mathfrak{S}$ of free states and therefore quantifies the particle correlation in the spirit of quantum chemistry.
Remarkably, the minimization in Eq.~\eqref{nonfreeness} can be explicitly performed. Given a state $\rho$, the minimizer is the unique free state $\Gamma_\rho\in\mathfrak{S}$ which has the same 1RDM as $\rho$~\cite{Gottlieb14}. As a result, the nonfreeness of $\rho$ can be determined  explicitly as~\cite{Gottlieb14}
\be
\mathcal{N}(\rho)= S(\gamma_\rho)+S(\mathbbm{1}-\gamma_\rho)-S(\rho),
\ee
where $S$ denotes the von Neumann entropy.

To make the connection between orbital correlations and nonfreeness/particle correlation precise, we now present a remarkable result that has not been properly acknowledged yet in quantum chemistry despite its potential far-reaching implications.
For this, we first consider the total \textit{orbital} correlation in a state obeying parity superselection rule (P-SSR as per \ref{subsec:SSR}), i.e., we refer to the \emph{finest} partition into single modes (associated with operators $a_i^{(\dagger)}$) of the one-particle Hilbert space. This means we quantify correlation with respect to $d$ subsystems given by the individual modes, see Sec. \ref{subsec:OrbCandE}.
Based on Eq.~\eqref{totcordsub}, it means that we need to compute the entropy $S(\rho_i)$ of every single mode reduced state $\rho_i$, obtained by tracing out all modes except the $i$-{th} one, that is $\rho_i=\Tr_{\backslash{i}}[\rho]$. Given that our total state $\rho$ is parity superselected, the reduced states takes the form $\rho_i=(1-p_i)\ket{0}\bra{0}+p_i\ket{1}\bra{1}$. Accordingly, the total orbital correlation follows as
\be \label{eq:IB}
\begin{split}
I_\mathcal{B}(\rho)=&\sum_i [-p_i\log_2{p_i}-(1-p_i)\log_2{(1-p_i)}]-S(\rho)\\
=& b((\gamma_\rho)_{d})-S(\rho)~.
\end{split}
\ee
The subscript $\mathcal{B}$ in \eqref{eq:IB} signifies that the value of total orbital correlation depends on the choice of one-particle basis $\mathcal{B}=\{\ket{a_i}\}$ underlying the operators $a_i^{(\dagger)}$. Moreover, in the second line, we have defined the binary entropy function $b(x)=-x\log_2{x}-(1-x)\log_2{(1-x)}$ and $(\gamma_\rho)_d$ denotes a diagonal matrix with entries equal to the diagonals of $\gamma_\rho$ in the particular basis $\{\ket{a_i}\}$. Now, one may ask to which value this total correlation can be reduced by choosing a different reference basis. To answer this question, one must minimize $I_\mathcal{B}$ over all one-particle bases $\mathcal{B}$ of $\mathcal{H}^{(1)}$. This finally leads to the remarkable result~\cite{gigena}
\be\label{orbpat}
\mathcal{N}(\rho)=\min_{\mathcal{B}}I_\mathcal{B}(\rho).
\ee
The proof of this equality uses tools from majorization theory ~\cite{Marshall1980InequalitiesTO,nielsen_majorization} to show that the minimising basis $\mathcal{B}$ is simply the one of the natural modes/spin-orbitals. In words, relation \eqref{orbpat} means nothing else than that the particle correlation measured through the nonfreeness is identical to the total orbital correlation minimized over all orbital reference bases $\mathcal{B}$. Accordingly, particle correlation corresponds to the minimal, thus intrinsic, complexity of many-electron wave functions while orbital correlation quantify their complexity relative to a fixed basis. Hence, the relation \eqref{orbpat} explains to which extent orbital optimization can reduce the computational complexity of an $N$-electron quantum state in quantum chemistry.

\subsection{Relation to Hartree-Fock Theory}\label{subsec:HF}
\begin{figure*}[htb]
\centering
\frame{\includegraphics[width=1.0\linewidth]{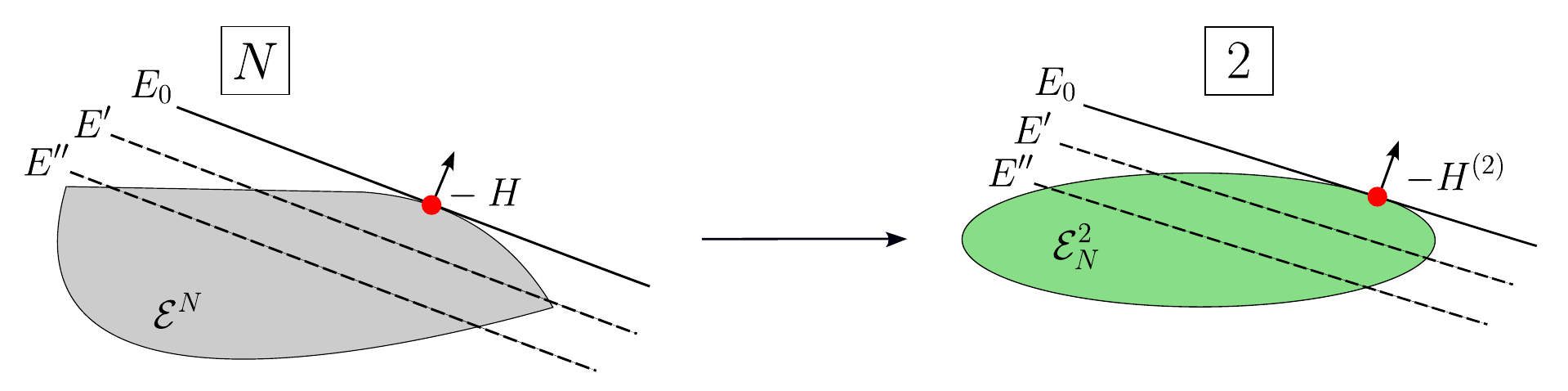}}
\caption{Illustration of the ground state search on the $N$-fermion Hilbert space (left). The dashed lines depict hyperplanes of constant energy $\Tr_N[H\rho]$ with normal vector $H$ which are shifted in direction of $-H$ until the boundary of $\mathcal{E}^N$ is touched. For Hamiltonians with at most two-body interactions $H^{(2)}$ (right), this reduces to a minimization over the set $\mathcal{E}^2_N$ of ensemble $N$-representable 2RDMs. Due to the linearity of $\mathrm{Tr}_2[H^{(2)}D]$ the minimum (red dot) is again attained at the boundary. \label{fig:Coulson}}
\end{figure*}

In this section, we first recall the Coulson challenge which emphasizes the significance of the 2RDM. Then, we present a remarkable connection between the 1RDM, 2RDM, free states and the nonfreeness. This provides additional evidence for the crucial role all these quantities should play in quantum chemistry.

Solving the ground state problem for large systems is cursed by the exponential scaling of the dimension of the $N$-fermion Hilbert space $\mathcal{H}^{(N)}$ \eqref{eq:Nfermionspace} with the system size. Most numerical methods to calculate the ground state energy are based on the Rayleigh-Ritz variational principle, which corresponds to an minimization of the expectation value $\bra{\Psi}H\ket{\Psi}$ over all $\ket{\Psi}\in \mathcal{H}^{(N)}$. This does not exploit, however, that most physical Hamiltonians include only pair-wise interactions. Based on this observation, Coulson formulated in the closing speech at the 1959 Boulder conference in Colorado the vision to replace the $N$-fermion wave function by the two-particle reduced density matrix (2RDM)
\begin{equation}
D := \begin{pmatrix}
N\\2
\end{pmatrix}\Tr_{N-2}[\Gamma]
\end{equation}
which contains considerably fewer degrees of freedom~\cite{Coulson60, Coleman01, Coleman-book}. In that case, the ground state search would simplify according to
\begin{eqnarray}\label{eq:RR-E0}
E_0 &=& \min_{\Gamma\in\mathcal{E}^N}\mathrm{Tr}_N[\Gamma H]\nonumber\\
&=&\min_{D\in \mathcal{E}^2_N}\Tr_2[DH^{(2)}]\,.
\end{eqnarray}
Here, in conceptual analogy to Eq.~\eqref{eq:E1N} for 1RDMs, we introduced the set $\mathcal{E}^2_N$ of ensemble $N$-representable 2RDMs, i.e., $\mathcal{E}^2_N$ consists of exactly those 2RDMs $D$ which are compatible with an ensemble $N$-fermion state~\cite{Coleman63, GP64, Kummer67, M12, hchain2023}. Moreover, we introduced the restriction $H^{(2)}$ of $H$ onto the two-particle level. However, working on the two-particle level does not trivialize the ground state problem: a significant part of the computational complexity of minimizing over an exponentially large $N$-fermion Hilbert space is shifted to the problem of finding an efficient description of the set $\mathcal{E}^2_N$ (Coulson challenge)~\cite{GP64, Kummer67, M12, hchain2023}. We further illustrate Eq.~\eqref{eq:RR-E0} in Fig.~\ref{fig:Coulson}.
Based on the Hilbert-Schmidt inner product, the expectation value $\mathrm{Tr}_N[\Gamma H] = \langle H, \Gamma\rangle_N$ describes a linear functional on the space of linear operators on $\mathcal{H}^{(N)}$. Then, the ground state energy $E_0$ is obtained by shifting the corresponding hyperplane of constant value in the direction of $-H$ until it touches the boundary (left side). Due to the linearity of the partial trace map, this linear structure immediately translates to the two-particle level illustrated on the right part of Fig.~\ref{fig:Coulson}. This illustration of Eq.~\eqref{eq:RR-E0} highlights the huge impact that the geometry of quantum states (recall also Sec.~\ref{subsec:hierarchy}) and tools from convex analysis have in quantum chemistry. In particular, these concepts were so effective that their straightforward application led recently to more comprehensive foundations of functional theories~\cite{Schilling18, LCLS21, Penz22, Helgaker22, LCS23}.

The minimization over $N$-fermion states in Eq.~\eqref{eq:RR-E0} can be relaxed to states on the Fock space $\F$ by introducing a chemical potential that fixes the total particle number $N$. For general states on $\F$ with indefinite particle number, their 2RDMs $D$ can be defined in a similar fashion as in Eq.~\eqref{eq:1RDM-defFock} by introducing a map $\mu^{(2)}$,
\begin{equation}\label{eq:2RDM-defFock}
\mu^{(2)}: \rho \mapsto D_\rho\,,\quad \mbox{with}\,(D_\rho)_{ij;kl}=\Tr[\rho f_l^\dagger f_k^\dagger f_i^{\phantom{\dagger}}f_j^{\phantom{\dagger}}]\,.
\end{equation}
Moreover, the ground state energy can be approximated by restricting the variational energy minimization to a submanifold of states. A quite crude but well-known example thereof is Hartree-Fock (HF) theory where the minimization is restricted to the manifold of Slater determinants. Since the 1RDMs of Slater determinants are idempotent, $\gamma^2=\gamma$, their 2RDMs are given by
\begin{equation}\label{eq:D-HF}
D_{\mathrm{HF}}(\gamma) = \frac{1}{2}(\mathbbm{1}-\mathbbm{Ex})\gamma\otimes \gamma\,.
\end{equation}
Thus, the HF energy $E_\mathrm{HF}$ for a Hamiltonian $H=h+W$, where $W$ denotes the two-body interaction, follows from minimizing the HF energy functional $\mathcal{E}_{\mathrm{HF}}(\gamma) = \Tr_1[h\gamma]+\Tr_2[WD_{\mathrm{HF}}(\gamma)]$ over all idempotent 1RDMs (we now skip the superscript ${(2)}$ in the Hamiltonian denoting its restriction to the two-particle level).
In his seminal work, Lieb showed that for positive semi-definite interactions $W\geq 0$ this minimization can be relaxed to the set of all ensemble $N$-representable 1RDMs $\gamma\in \mathcal{E}^1_N$~\cite{LiebHF}.
By exploiting the one-to-one relation between 1RDMs and free states, it further follows that a relaxation from Slater determinant states on $\mathcal{H}^{(N)}$ to free states on the Fock space does not alter the outcome of the energy minimization for $W\geq 0$~\cite{BLS94}. This result exploits the remarkable fact that the 2RDM of \textit{any} free state is given by precisely $D_{\mathrm{HF}}$, that is
\begin{equation}
\mu^{(2)}(\Gamma) = D_{\mathrm{HF}}(\gamma_\Gamma)\quad\forall\,\Gamma\in\mathfrak{S}\,.
\end{equation}
Thus, despite $D_{\mathrm{HF}}(\gamma)$ being not ensemble $N$-representable for $\gamma^2\neq\gamma$, it is indeed representable to a free state on the Fock space for all $\gamma\in \mathcal{E}^1_N$. Moreover, the corresponding free state is simply given by the free state \eqref{eq:1RDMtofree} mapping to the 1RDM $\gamma$. In fact, the so-called Lieb variational principle ~\cite{LiebHF, BBKM14, Helgaker22} and resulting HF functional $F_\mathrm{HF}(\gamma) := \Tr_2[WD_{\mathrm{HF}}(\gamma)]$ provided the foundation for more sophisticated functional approximations in one-particle reduced density matrix functional theory~\cite{M84, BB02, GPB05, PG16, P21, LCS23}.

\begin{figure*}[htb]
\centering
\frame{\includegraphics[width=1.0\linewidth]{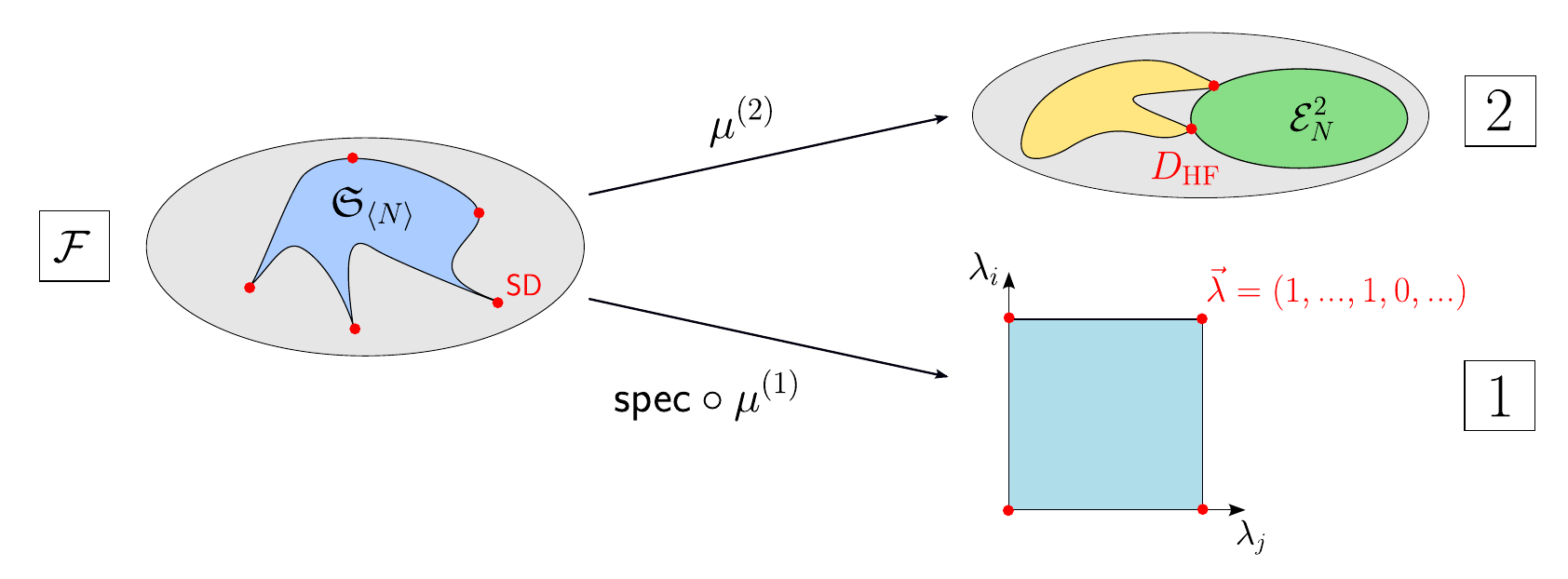}}
\caption{Illustration of the intersection of $\mathfrak{S}$ with a hyperplane of fixed average particle number $N$ (see also Fig.~\ref{fig:freestates-1RDM}) denoted by $\mathfrak{S}_{\langle N\rangle}$  and its image under the maps $\mu^{(1/2)}$ leading to the respective  1RDMs and 2RDMs. The red dots (left) illustrate the Slater determinants. The image of $\mathfrak{S}_{\langle N\rangle}$ (light blue, left) under $\mu^{(2)}$ (yellow set, right) only intersects with the set of ensemble $N$-representable 2RDMs (green) at those 2RDMs whose preimages are Slater determinants. On the one-particle level, we illustrate the Pauli hypercube (blue) of admissible natural occupation number vectors $\vec{\lambda}$ (see text for more details).
\label{fig:Freestates-2RDM-1RDM}}
\end{figure*}

We illustrate the relation between free states and their 2RDMs and 1RDMs in Fig.~\ref{fig:Freestates-2RDM-1RDM}. The set $\mathfrak{S}_{\langle N\rangle}$ depicts the intersection of the set of free states $\mathfrak{S}$ with the hyperplane of constant average particle number $\langle \hat{N}\rangle_{\Gamma}=N$ as explained in Sec.~\ref{sec:free-states}. This is meaningful since also on the two and one-particle level we refer to a fixed particle number in Fig.~\ref{fig:Freestates-2RDM-1RDM}. The set of all states on the Fock space $\F$ is illustrated in gray. In contrast to $\mathfrak{S}$ in Fig.~\ref{fig:freestates-1RDM}, the intersection with a hyperplane might have the effect that not all extremal elements of $\mathfrak{S}_{\langle N\rangle}$ are Slater determinants (red dots) anymore. We then depict the image of the set $\mathfrak{S}_{\langle N\rangle}$ under the map $\mu^{(2)}$ on the two-fermion level by the yellow set. As explained above, this set only intersects with the set $\mathcal{E}^2_N$ of ensemble $N$-representable 2RDMs at those 2RDMs whose preimage is a Slater determinant (illustrated by red dots on the right side). On the one-particle level, we compose the map $\mu^{(1)}$ introduced above with the spectral map $\mathrm{spec}(\cdot)$ which maps a 1RDM to its vector $\vec{\lambda}$ of eigenvalues $\lambda_i$. Due to the one-to-one relation between free states and 1RDMs $\gamma\in \mathcal{E}^1_N$ the image under this composed map is given by the Pauli hypercube (blue) whose vertices correspond to idempotent 1RDMs with eigenvalues $\lambda_i\in\{0,1\}$ (red dots).

\section{Examples and Illustrations}\label{sec:examples}

In the previous sections, we have established the notions of particle and orbital correlation. In particular, we presented a fundamental relation between them: the particle correlation measure of nonfreeness~\eqref{nonfreeness} equals the orbital-minimized total orbital correlation (see Eq.~\eqref{orbpat}). In this section, we shall demonstrate this link between the orbital and particle picture, with both analytic and numerical examples. Moreover, we will relate these quantities with the so-called configuration interaction (CI) entropy of wave functions, a direct but computationally costly quantifier of the multireference character.

\begin{figure*}[htb]
\centering
\frame{\includegraphics[width=1.0\linewidth]{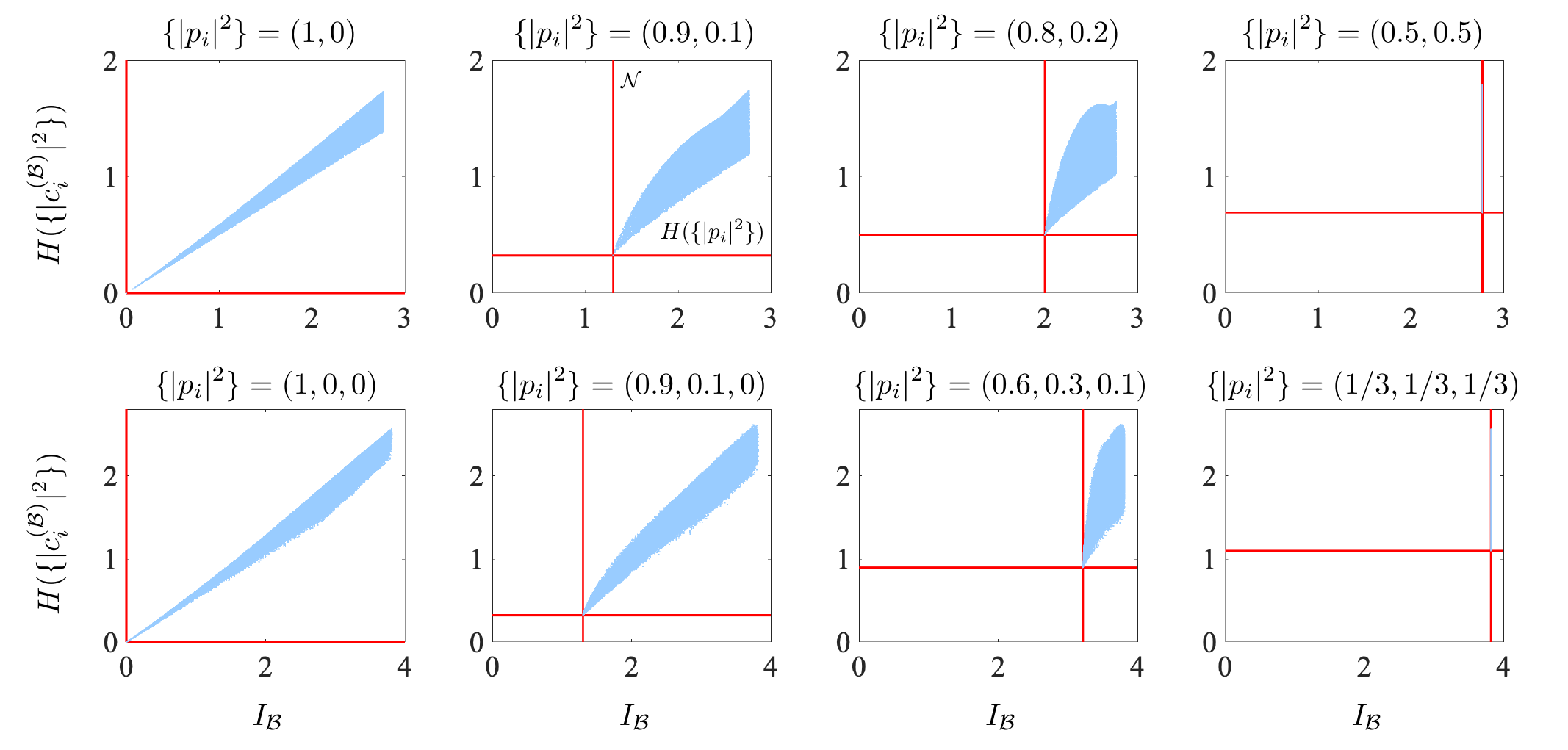}}
\caption{Relation between total orbital correlation $I_\mathcal{B}(|\Psi_{2e}\rangle\langle\Psi_{2e}|)$ and Shannon entropy $H(\{|c_i^{(\mathcal{B})}|^2\})$ of the CI expansion coefficients of two-electron quantum state \eqref{eqn:2el_no} in $2K$ modes for numerous randomly sampled orbital reference bases $\mathcal{B}$. The horizontal and vertical red lines indicate the minima of the two quantities, given by $H(\{|p_i|^2\})$ and $\mathcal{N}$ (nonfreeness), respectively. In the first row we set $K=2$, and in the second row $K=3$. Different plots in the same row correspond to different choices of the parameters $\{|p_i|^2\}_{i=1}^{K}$ in \eqref{eqn:2el_no} that uniquely determine the multireference structure of the state.}\label{fig:2el_sampling}
\end{figure*}

\begin{figure*}[htb]
\centering
\frame{\includegraphics[width=1.0\linewidth]{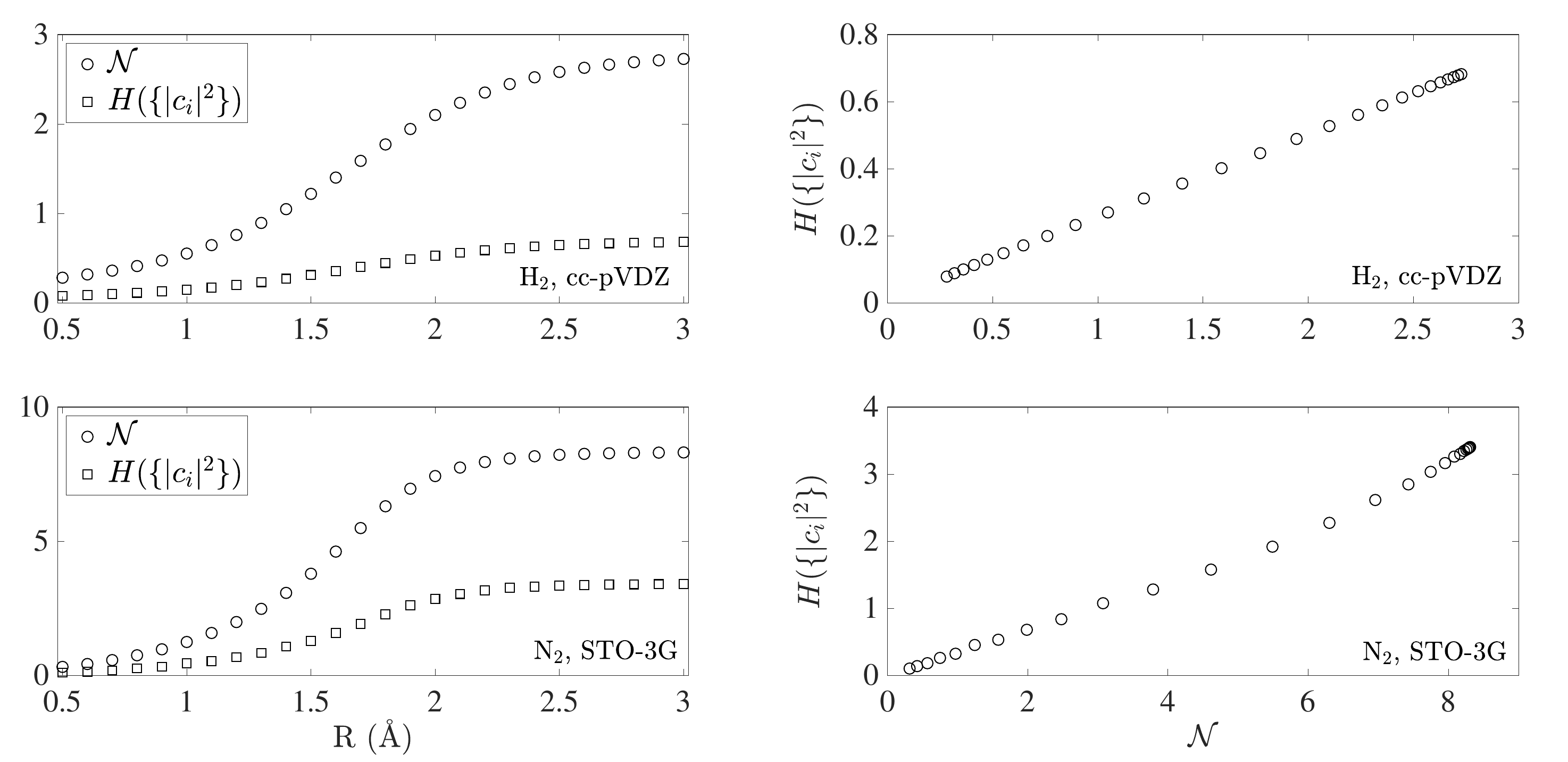}}
\caption{Nonfreeness $\mathcal{N}$ and Shannon entropy $H(\{|c_i|^2\})$ of the CI expansion coefficients in natural orbital basis for the Full CI ground state of $\mathrm{N_2}$ in the STO-3G basis at various internuclear distances $R$ (left), and $H(\{|c_i|^2\})$ as a function of $\mathcal{N}$ (right).}\label{fig:h2n2}
\end{figure*}

\subsection{Analytical example}\label{subsec:Exanaly}

For two electrons in $2K$ modes (spin-orbitals which form an orthonormal basis $\mathcal{B}$ of $\Ho$), a general quantum state can be expanded in terms of Slater determinants,
\begin{equation}
\begin{split}
    |\Psi_\mathrm{2e}\rangle = \sum_{1\leq j<k\leq 2K} c^{(\mathcal{B})}_{jk} f^\dagger_j f^\dagger_k |0\rangle = \sum_{i=1}^{{2K}\choose{2}} c^{(\mathcal{B})}_i |\mathrm{SD}_i\rangle \; .
\end{split}
\end{equation}
In the last equation of this configuration interaction (CI) expansion we collected the CI coefficients $c^{(\mathcal{B})}_{jk}$ into a vector $\vec{c}= \big(c^{(\mathcal{B})}_i\big)$. In the natural orbital basis, $|\Psi_\mathrm{2e}\rangle$ admits a compact form containing at most $K$ Slater determinants \cite{schliemann01fermion},
\begin{eqnarray}
    |\Psi_\mathrm{2e}\rangle = \sum_{i=1}^{K} p_i a^\dagger_{2i-1}a^\dagger_{2i}|0\rangle,
    \label{eqn:2el_no}
\end{eqnarray}
where we have reserved the operators $a^{(\dagger)}_{i}$ for the natural spin-orbitals or modes. From Sec.~\ref{sec:free-states}, it is established that precisely in this basis of modes, the total orbital correlation $I_{\mathcal{B}}(|\Psi_\mathrm{2e}\rangle\langle\Psi_\mathrm{2e}|)$ is at its minimum. In fact, the wave function $|\Psi_\mathrm{2e}\rangle$ is also expected to have the most compact form in the natural basis, where the compactness of the expansion is measured by the Shannon entropy $H(\{|c^{(\mathcal{B})}_i|^2\})$ of the squared CI coefficients $\{|c^{(\mathcal{B})}_i|^2\}$. We shall refer to it as the CI entropy. In other words, we conjecture for $N=2$ fermions that the CI entropy in any basis $\mathcal{B}$ satisfies
\begin{equation}
    H(\{p_i^2\}) \leq H(\{|c^{(\mathcal{B})}_i|^2\}), \quad \forall \mathcal{B}. \label{eqn:conj}
\end{equation}

To verify this conjecture we sampled various one-particle bases $\mathcal{B}$. To be more specific, for the case $K=2$ (the first row in Fig.~\ref{fig:2el_sampling}), $10^5$ $4\times4$ orthogonal matrices were sampled uniformly from the orthogonal group, which transform the natural orbitals to a target basis. For the case $K=3$ (the second row in Fig.~\ref{fig:2el_sampling}), $10^6$ $6\times6$ orthogonal matrices were sampled from the orthogonal group, and $10^6$ additional ones are sampled around the identity. In Fig.~\ref{fig:2el_sampling}, we present both quantities $I_{\mathcal{B}}(|\Psi_\mathrm{2e}\rangle\langle\Psi_\mathrm{2e}|)$ and $H(\{|c^{(\mathcal{B})}_i|^2\})$, for various states with different levels of \textit{intrinsic} multireference character modulated by the choice of parameters $p_i$'s in Eq. \eqref{eqn:2el_no} (i.e., the natural occupation numbers, up to a square). When only one $|p_i|^2$ is nonzero and therefore equal to 1, $|\Psi_\mathrm{2e}\rangle$ is a single Slater determinant. When various $|p_i|^2$'s are fractional, $|\Psi_\mathrm{2e}\rangle$ always contains some multireference character in any orbital basis.

First, we observe that indeed for each two electron state $|\Psi_\mathrm{2e}\rangle$, the two correlation quantities $I_\mathcal{B}(|\Psi_{2e}\rangle\langle\Psi_{2e}|)$ and $H(\{|c_i^{(\mathcal{B})}|^2\})$ are simultaneously minimized, specifically by the natural orbital basis. This provides the first evidence for our conjecture \eqref{eqn:conj}. In particular, when only two $|p_i|^2$'s are nonzero (which is the general case if the number of modes is $2K=4$), the minima of the two quantities can be shown to be related by
\begin{equation}
\begin{split}
    \min_\mathcal{B} I_\mathcal{B}(|\Psi_{2e}\rangle\langle\Psi_{2e}|) = 4 \min_\mathcal{B }H(\{|c_i^{(\mathcal{B})}|^2\}).
\end{split}
\end{equation}
Second, it is clear from the plots that when $I_\mathcal{B}(|\Psi_{2e}\rangle\langle\Psi_{2e}|)$ reduces, both the upper and lower bounds of $H(\{|c_i^{(\mathcal{B})}|^2\})$ are also reduced. Specially, when $I_\mathcal{B}(|\Psi_{2e}\rangle\langle\Psi_{2e}|)$ approaches its minimum, the gap between the two bounds of $H(\{|c_i^{(\mathcal{B})}|^2\})$ closes. Third, as the $|p_i|^2$'s get close to each other and the intrinsic multireference character of the state thus increases, the range of the values of $I_\mathcal{B}(|\Psi_{2e}\rangle\langle\Psi_{2e}|)$ shrinks. This range collapses to a point when the natural occupation numbers become identical (which is evidenced by the line of blue dots sitting right on top of the vertical red line in the last column in Fig.~\ref{fig:2el_sampling}). A similar effect is observed for $H(\{|c_i^{(\mathcal{B})}|^2\})$, but much less pronounced. Finally, if we focus on the natural orbitals (represented by the data point at the intersection of the two red lines), we see that the CI entropy is monotonic with the total spin-orbital correlation.

We remark that the conjectured inequality \eqref{eqn:conj} can be proven analytically for any pure state of two fermions ($N=2$). While a detailed proof goes beyond the scope of this article, in~\cite{Vedralwithproof} the interested reader can find the proof of a similar inequality which arises in the evaluation of the quantum correlation (Eq. \eqref{qcorr}) of any pure state of two distinguishable particles.

\subsection{Numerical illustrations}\label{subsec:Exnum}

To elaborate on our observations from the two-electron examples, we now inspect the relation between the nonfreeness $\mathcal{N}$ \ref{nonfreeness} and the CI entropy for the ground state of $\mathrm{H_2}$ (cc-pVDZ, 2 electrons in 10 orbitals) and $\mathrm{N_2}$ (STO-3G, 14 electrons in 10 orbitals) in the natural basis. For each molecule, we calculate first the full CI ground state based on the previously obtained HF orbitals. Then, we obtain the natural spin-orbitals by diagonalizing the 1RDM $\gamma$ of the ground state. Lastly, we perform another full CI calculation based on the natural spin-orbitals, and acquire the CI coefficient vector $\vec{c}$ in this basis. This procedure is done using the PySCF package~\cite{sun2018pyscf}, and is repeated for various internuclear distances of the molecules, ranging from $0.5$\AA\hspace{0.2mm} to $3$\AA.

We present the results in Fig.~\ref{fig:h2n2}. For both molecules, as the internuclear distance $R$ increases from below equilibrium, the multireference character of the ground states also increases. This can be directly seen from the left column, where both the nonfreeness $\mathcal{N}$ and the CI entropy $H(\{|c_i^{(\mathcal{B})}|^2\})$ grow with the internuclear distances. Moreover, the relation between the nonfreeness $\mathcal{N}$ and the CI entropy $H(\{|c_i^{(\mathcal{B})}|^2\})$ is monotonic (and even almost linear for $\mathrm{H}_2$), as the second column in Fig.~\ref{fig:h2n2} clearly demonstrates. This again highlights the high potential of the easily accessible nonfreeness as a universal tool for characterizing multireference wave functions, in place of the cumbersome if not inaccessible CI entropy $H(\{|c_i^{(\mathcal{B})}|^2\})$.

Our numerical results confirm that some of the insights from the analytic two-electron examples indeed extend to larger systems. Both the minimized total spin-orbital correlation, i.e., the nonfreeness $\mathcal{N}$, and the CI entropy $H(\{|c_i|^2\})$ in the natural basis are valid quantitative descriptors of the multireference character of the ground state. More importantly, the two descriptors are found to be monotonic functions of each other for both molecules. This suggests that the simple nonfreeness $\mathcal{N}$, which only involves the 1RDM $\gamma$, can reveal the high complexity of the wave function encoded in the CI expansion equally well as the CI entropy which is difficult to calculate in practice.

\section{Summery and Conclusions}\label{sec:conclu}

In order to foster synergy between quantum chemistry and quantum information theory, we translated the concepts of entanglement and correlation into the context of quantum chemical systems. By exploiting the formalism of first and second quantization we established two conceptually distinct notions of correlation in fermion systems. To be more specific, we first recalled that subsets of orbitals define in a precise way quantum subsystems by referring to their respective algebras. This in turn then allowed us to apply the common formalism of `local operations and classical communication' (LOCC) to introduce a notion of orbital correlation and entanglement. In particular, to make it operationally meaningful, we elucidated why and how the fundamental number parity superselection rule needs to be taken into account. Moreover, to invite the quantum chemists to join the ongoing second quantum revolution, we explained for which quantum information processing tasks the corresponding orbital entanglement could be used for. Quite to the contrary, electrons themselves do not obey the axioms of quantum subsystems and thus the paradigm of LOCC cannot be applied. Instead, we thus defined the ground and thermal states of noninteracting electron systems as the particle uncorrelated states. 
Measuring then the minimal distance of a quantum state $\rho$ through the quantum relative entropy to the manifold of those `free states', results directly in a measure of particle correlation~\cite{Gottlieb_2005,Gottlieb14}. It is given by the particle-hole symmetrized von Neumann entropy $S$ of the corresponding 1RDM $\gamma_\rho$ modified by the entropy of the total state, i.e., $\mathcal{N}(\rho)= S(\gamma_\rho)+S(\mathbbm{1}-\gamma_\rho)-S(\rho)$.

We then demonstrated that particle correlation equals total orbital correlation minimized with respect to all orbital reference bases. Accordingly, particle correlation equals the minimal, thus intrinsic, complexity of many-electron wave functions while orbital correlation quantifies their complexity relative to a basis. From a practical point of view, the particle correlation therefore defines the correlation threshold to which orbital optimization schemes can reduce the representational complexity of the many-electron wave function. Prime examples for methods with a particularly strong dependence on the orbital reference basis are the density matrix renormalization group (DMRG)-ansatz, as well as, variational quantum eigensolvers (VQE) in quantum computing. To further strengthen the connection between the particle and orbital picture we presented
an inherent relation between free states and Hartree-Fock theory: the states which were defined as particle uncorrelated from a quantum information perspective are precisely those which underlie the construction of the pivotal Hartree-Fock functional in one-particle reduced density matrix functional theory. Accordingly, it can be expected that the measure of particle correlation might be connected to the two-electron cumulant which is discarded in Hartree-Fock theory.

We illustrated all these concepts and analytical findings in few-electron systems.
With an analytic example of two-electron states and a numerical one concerning two concrete molecules, we made two instructive observations: (i) (At least for states of two electrons) both the total spin orbital correlation and the entropy of the CI coefficients are minimized in the natural orbital basis. (ii) The particle correlation, which is the minimized total spin orbital correlation, is a good approximation (up to a factor) to the complicated CI entropy relative to the natural orbitals, which measures directly the multireference character of the wave function expansion.
At the same time, these results suggested a general guiding principle for simplifying the structure of the wave function: By reducing the total (spin) orbital correlation, one effectively trims away the excessive complexity in the wave function due to a sub-optimal orbital representation.

In summary, we believe that the fermion-compatible correlation and entanglement measures presented in this work facilitate a complete characterization and successful exploitation of the quantum resourcefulness of atoms and molecules for information processing tasks. In return, the profound connection between orbital and particle correlation, as well as their role in evaluating the multireference character of wave functions, could stimulate developments of novel and more efficient approaches to the electron correlation problem.

\begin{acknowledgements}
We acknowledge financial support from the German Research Foundation (Grant SCHI 1476/1-1), the
Munich Center for Quantum Science and Technology,
and the Munich Quantum Valley, which is supported by
the Bavarian state government with funds from the Hightech Agenda Bayern Plus. J.L. acknowledges funding from the International Max Planck Research School for Quantum Science and Technology (IMPRS-QST).
\end{acknowledgements}

\bibliography{faraday}

\end{document}